# A CryptoCubic Protocol for Hacker-Proof Off-Chain Bitcoin Transactions

## Leonard Apeltsin

**1 INTRODUCTION**

Off-Chain transactions allow for the immediate transfer of Cryptocurrency between two parties, without delays or unavoidable transaction fees. Such capabilities are critical for mainstream Cryptocurrency adaption. They allow for the "Coffee-Coin Criteria"; under which a customer orders a coffee and pays for that coffee in bitcoins. This is not possible with On-Chain transactions today. No customer is willing to wait for 20 minutes for their coffee transaction to receive six public confirmations. Neither will the customer pay a 20 cent transaction fee when the coffee costs approximately $2. Only a quick and free Off-Chain transaction will satisfy our coffee-guzzling consumer. Otherwise, the customer will stick to credit cards and cash, and Bitcoin will face a limited future with regards to everyday use.

Unfortunately, all existing Off-Chain transaction protocols are notoriously unreliable [https://en.bitcoin.it/wiki/Off-Chain_Transactions#Trusted_Third_Parties]. The current generation of third-party facilitators are vulnerable to hacker-based attacks. As Mt. Gox tragically demonstrated, centralized-transaction institutions are easy targets for Cryptocurrency thieves. The slightest security flaw in a third-party system will pounced on by hackers, who will proceed to devour it like ants devouring a crab. Furthermore, the possible issue of fraud is constantly paramount; third-party facilitators offer little proof that they actually hold the bitcoins of their clients [https://en.bitcoin.it/wiki/Off-Chain_Transactions#Auditing]. Under such circumstances, it no wonder that the Public treats most Cryptocurrency services with a constant shadow of suspicion.

Fortunately for users, trusted third-parties do exist; Coinbase is one such company. Coinbase takes a series of concrete and careful steps to circumvent the weak points in its protocol. These steps include employee background checks, decentralized cold storage in bank vaults scattered across the globe, and a team of security experts on-hand to mitigate against cyber-attack [https://coinbase.com/security]. With a multi-million dollar effort, Coinbase was able to construct a massive stronghold in order to defend its vulnerable interior from the barbaric hacker hordes. Given these resources, Coinbase is well on its way to becoming the "Gmail" of the Cryptocurrency community [http://www.coindesk.com/coinbase-gmail-bitcoin/].

The onset of Coinbase has brought benefits to amateur Bitcoin users. Entrepreneurial developers on the other hand, have not fared nearly as well. Many organizations lack the financial resources required to run secure Off-Chain transactions. An entrepreneur struggling to design his own Bitcoin coffee-purchasing app will be unable to safely do so on his own. He must rely on the Coinbase transfer API, with all its licensed restrictions and limitations [https://coinbase.com/legal/user_agreement]. This centralized reliance on a single third-party provider severely reduces the freedoms of the Bitcoin developer community. Furthermore, such vast centralization poses a singular threat to the Bitcoin paradigm as a whole. Suppose a hacker finds a crack in Coinbase-armor; all Off-Chain accounts will then be compromised! That hack would undermine all Coinbase-dependent tools and apps, thereby obliterating the trust of the casual Cryptocurrency enthusiast.

For Bitcoin to flourish, its anti-hierarchy principles must be applied to safe Off-Chain transactions. First and foremost, we need a new HACKER-PROOF protocol that can easily be executed by any experienced developer. Preferably, the protocol will be open-sourced for full reliability and transparency. The developer community must come to the consensus that the protocol itself remains unbreakable.

In this paper we present one such solution; the CryptoCubic (CC) Protocol. The CC Protocol employs MultiSig technology to safely transfer ownership of actual bitcoin addresses between individual users. In the following documentation we will describe in iterative detail the science behind our CryptoCubic technique, as well as all its individual steps. We will also discuss all possible theft-driven attacks against our system, and the how protocol intrinsically defends itself against such BlackHat exploitations. It is our aim to meticulously show how the CryptoCubic Protocol is hacker-proof in all significant ways.

**2 DEFINING A HACKER-PROOF PROTOCOL**

How does one rigorously demonstrate that a protocol is actually hacker-poof? Any non-trivial system may potentially be manipulated in a seemingly infinite number of permutations. Therefore, in order to confidently claim that a protocol is safe, one must first define the settings under which that safety is guaranteed. One must lay down the rules of the game, so to speak, in order to then demonstrate that these rules are inherently unbreakable.

We begin to define the rules through a series of simple assumptions. First and foremost, we assume that the server is benign, and not malicious. The server does not purposefully attempt to cheat or manipulate its users. Second of all, we assume that the server executes the protocol exactly as specified; no unauthorized alterations or modifications are allowed. Furthermore, we assume the presence of a security apparatus capable of immediately detecting any deviations from the protocol. That is, if a malicious agent gains server access and modifies the code, then protocol execution will immediately terminate until the intrusion is resolved. Finally, we assume any calculated variables occurring within an running program are wholly inaccessible outside the boundary of that program. For example, if an on-server program dynamical computes some secret variable X, then that variable shall remain hidden from the world until it is specifically outputted to a memory-location on the server.

Additionally, let us consider one other significant supposition. We conjecture it is possible to design a simple so-called "self-destructive" storage mechanism that is 100% secure. What is a self-destructive storage mechanism? Imagine a database table whose content is limited to a certain pre-specified Source X. Output requests to the rows in that table may come from multiple unspecified sources. We may ping a particular row from the table to check if its empty or not. However, accessing the data in that row leads to the data's immediate deletion. Afterwards, the data may not be replaced without the deliberate permission of Source X. Let us consider an actual example; X produces Y, and stores Y in a self-destructive database. That action is represented as X-- [Y]. We ping the database to determine that Y is indeed present, though we do not know its contents. Next, we input a retrieval request for Y in order to obtain its true identity . Y is immediately retrieved, and is automatically deleted from the database. We may try to manipulate the system in order to subtly put Y back its original location, but we will not succeed without the direct permission of resource X. That, in a nutshell, is the function of a self-destructive database. Though its maximal level of proven security remains to be determined, we venture to conjecture that a totally secure self-destructive mechanism is mathematically possible.

Based on the above-stated assumptions, the server acts as a trusted third-party to its users. The server is also a potential target for malicious hackers and thieves. In fact, we shall directly associate hacking with theft. Hackers will willfully attack the server for the purpose of financial gain. Under such conditions, wanton and profit-less destruction is not considering hacking. Deletion of a server's contents is not a hack. Neither is smashing that server with a sledgehammer. We define such destructive activities as acts of vandalism. Defending against vandalism requires a secure system of storage and backup, which will not be discussed in this paper. We will focus instead on for-profit attacks. Our concern is the thief breaking into the vault, not the arsonist trying to burn down the bank.

Thus we define a hacking attack as a deliberate attempt to manipulate the protocol for the purpose of illicit financial gain. A successful hack entails that an attacker illicitly obtains all necessary crypo-keys needed to execute an On-Chain Bitcoin transaction. These keys may be obtained in a variety of ways; ranging from the direct replication of server-stored data to the more subtle counterfeit emulation of protocol-specific signals. A hacker-proof protocol must successfully defend against all these myriad attacks. We shall hence develop one such protocol in the subsequent sections of our paper.

## 3  MULTISIG OFF-CHAIN TRANSACTIONS

In a standard Bitcoin transaction, a single private key is required to transfer funds from User_A to User_B. Whoever holds that key controls the funds, therby making it a dangerous single-point target for digital attacks. The onset of Bitcoin-based MultiSig cryptography greatly helps alleviate that threat. Let us a consider simple 2-of-2 MultiSig system. Two unique private keys, Sig_U and Sig_S, are associated with a single public address ADD. Both keys are required to control the funds within that address. Illicitly obtaining one but not the other private key is not enough to instigate the hack.

Suppose that we instigate a new relationship between User_A and Server_S. The Server then creates a 2-of-2 MultiSig key-pair associated with an address ADD. The key-pair and the address exist within a dynamic running process. They have not yet been stored in server memory, and are not accessible to predatory hackers, based on our predefined criteria. We represent this transitory state using notation <Sig_U,Sig_S,ADD>. At this point in our execution, the memory states of User_A and Server_S exist as follows:

| USER_A | SERVER_S |
| --- | --- |
|  | <Sig_U,Sig_S,ADD> |

Next, Server_S establishes a secure connection with User_A. Sig_U and ADD are transferred over to User_A, to be stored in his protocol client's memory.

| USER_A | SERVER_S |
| --- | --- |
| Sig_U | <Sig_U,Sig_S,ADD> |
| ADD |  |

Afterwards, Sig_S is transferred to the memory of the Server. Given the sensitive nature of Sig_S, we choose to treat its storage very careful. As a result, we load Sig_S into a self-destructive database; of the sort that is discussed in Section 2. That transfer is represented as <Sig_U,Sig_S,ADD> -- [Sig_S], where <Sig_U,Sig_S,ADD> is a permited self-destructive database input source.

| USER_A | SERVER_S |
| --- | --- |
| Sig_U | <Sig_U,Sig_S,ADD> -- [Sig_S] |
| ADD |  |

Finally, the dynamic procedure containing the variables <Sig_U,Sig_S,ADD> reaches termination. The temporary variables cease to exist in any form within the Server. Server_S is left completely unaware of the contents of private key Sig_U.

| USER_A | SERVER_S |
|---|---|
| Sig_U | [Sig_S] |
| ADD | |

At this point, User_A executes an On-Chain Bitcoin transaction from an exterior wallet, thereby transferring $10 to address ADD.

| USER_A | SERVER_S |
|---|---|
| Sig_U | [Sig_S] |
| ADD ($10) | |

Now User_A encounters User_B, who has no connection to the Server.

| USER_A | SERVER_S | USER_B |
|---|---|---|
| Sig_U | [Sig_S] | |
| ADD ($10) | | |

User_A and User_B initialize a data-exchange transaction, where User_B receives both ADD ($10) and Sig_U from User_A.

| USER_A | SERVER_S | USER_B |
|---|---|---|
| Sig_U | [Sig_S] | Sig_U |
| ADD ($10) | | ADD ($10) |

At this juncture, User_B has multiple options. He can transfer Sig_U back to User_A. He can transfer Sig_U to some other User_C. Finally, he can request Sig_S directly from the Server, resulting in its immediate deletion from the self-destructive database.

| USER_A | SERVER_S | USER_B |
|---|---|---|
| Sig_U | | Sig_U |
| ADD ($10) | | Sig_S |
| | | ADD ($10) |

Once User_B obtains Sig_S, he will gain instantaneous control of the $10 in address ADD. What we have just described is MultiSig Off-Chain transaction. Of course, the aforementioned transaction is exceedingly insecure. There are many reasons for this, but the foremost cause of insecurity is the unreliability of User_A. What if User_A grabs Sig_S from Server_S after the

transaction is completed? What if User_A posted the value of Sig_U on some shady Darknet hacker forum? What if User_A is just one of many previous Off-Chain Sig_U recipients, any of which could have comprised its contents? User_B remains consistently aware that one or more Sig_U-possessing individuals could in theory hack the Server, thereby stealing all his funds. This is unacceptable; we must employ cryptography to make our MultiSig transactions more secure.

## 4 ENCRYPTED MULTISIG OFF-CHAIN TRANSACTIONS

Let us consider the following MultiSig encrypted schema; Server_S interacts with User_A. User_A immediately produces an asymmetrical public/private pair of keys; (Ka,Ka_Public). Ka_Public can encrypt a string that may only be decrypted using Ka.

| USER_A | SERVER_S |
| --- | --- |
| Ka | |
| Ka_Public | |

User_A transfers the public key to Server_S. The Server then creates a symmetric key Ks, which can be used to both encrypt and decrypt data.

| USER_A | SERVER_S |
| --- | --- |
| Ka | Ks |
| Ka_Public | Ka_Public |

Server_S proceeds to input both keys into a dynamic procedure that is not accessible from memory.

| USER_A | SERVER_S |
| --- | --- |
| Ka | <Ks,Ka_Public> |
| Ka_Public | Ks |
| | Ka_Public |

The dynamic procedure generates three MultiSig components; Sig_U, Sig_S, and ADD.

| USER_A | SERVER_S |
| --- | --- |
| Ka | <Ks,Ka_Public,Sig_U,Sig_S,ADD> |
| Ka_Public | Ks |
| | Ka_Public |

The dynamic procedure encrypts Sig_U using Ka_Public, in order to create cypher Ea. In addition, the procedure encrypts Sig_S using Ks, in order to produce cypher Es.

| USER_A | SERVER_S |
|---|---|
| Ka | <Ks,Ka_Public,Sig_U,Sig_S,ADD,Ea,Es> |
| Ka_Public | Ks |
|  | Ka_Public |

The running procedure on Server_S securely transfers Es and ADD to User_A.

| USER_A | SERVER_S |
|---|---|
| Ka | <Ks,Ka_Public,Sig_U,Sig_S,ADD,Ea,Es> |
| Es | Ks |
| ADD | Ka_Public |
| Ka_Public |  |

The running procedure on Server_S transfers Ea into a self-destructive database

| USER_A | SERVER_S |
|---|---|
| Ka | <Ks,Ka_Public,Sig_U,Sig_S,ADD,Ea,Es> -- [Ea] |
| Es | Ks |
| ADD | Ka_Public |
| Ka_Public |  |

The running procedure finally terminates, thereby destroying all its inaccessible contents.

| USER_A | SERVER_S |
|---|---|
| Ka | [Ea] |
| Es | Ks |
| ADD | Ka_Public |
| Ka_Public |  |

The combined data contents of User_A and Server_S form a "CryptoSquare", which is highlighted in following table.

| USER_A | SERVER_S |
|---|---|
| Ka | [Ea] |
| Es | Ks |
| ADD | Ka_Public |
| Ka_Public |  |

The CryptoSquare ensures that neither User_A nor Server_S are able to obtain either of the unecrypted signature-keys (Sig_U,Sig_S) without mutual collaboration. User_A retains no knowledge of these variables. In order to make transfers from address ADD, User_A needs to request both cypher Ea and key Ks from Server_S, which will result in the immediate deletion of [Ea] within the server's self-destructive database.

The significance of the CryptoSquare first becomes apparent when User_A encounters User_B, after adding funds to address ADD.

| USER_A | SERVER_S | USER_B |
| --- | --- | --- |
| Ka | [Ea] | |
| Es | Ks | |
| ADD ($10) | Ka_Public | |
| Ka_Public | | |

User_A initiates an encrypted Off-Chain transaction by transferring Es and ADD ($10) to User_B

| USER_A | SERVER_S | USER_B |
| --- | --- | --- |
| Ka | [Ea] | Es |
| Es | Ks | ADD ($10) |
| ADD ($10) | Ka_Public | |
| Ka_Public | | |

User_B responds by creating the asymmetric key-pair (Kb, Kb_Public).

| USER_A | SERVER_S | USER_B |
| --- | --- | --- |
| Ka | [Ea] | Kb |
| Es | Ks | Es |
| ADD ($10) | Ka_Public | Kb_Public |
| Ka_Public | | ADD ($10) |

User_B transfers Kb_Public to Server_S.

| USER_A | SERVER_S | USER_B |
| --- | --- | --- |
| Ka | [Ea] | Kb |
| Es | Ks | Es |
| ADD ($10) | Kb_Public | Kb_Public |
| Ka_Public | Ka_Public | ADD ($10) |

The Server requests and receives permission from User_A to initiate the transaction. Afterwards, the server initializes a dynamic procedure that accesses Ea from the self-destructive database. Cypher Ea is eliminated from memory; it now exists solely within the inaccessible dynamic procedure.

| USER_A | SERVER_S | USER_B |
|---|---|---|
| Ka | <Ea> | Kb |
| Es | Ks | Es |
| ADD ($10) | Kb_Public | ADD ($10) |
| Ka_Public | Ka_Public | Kb_Public |

Server_S requests a copy of key Ka from User_A. It then confirms the key Ka is a proper match for Ka_Public. If Ka is not received, or a proper confirmation is not made, then Ea will once again be placed into a self-destructive database, and the transaction will be determined. Otherwise, the transaction will continue.

| USER_A | SERVER_S | USER_B |
|---|---|---|
| Ka | <Ea> | Kb |
| Es | Ks | Es |
| ADD ($10) | Kb_Public | ADD ($10) |
| Ka_Public | Ka | Kb_Public |
|  | Ka_Public |  |

The dynamic procedure on Server_S loads the value of Ka from server memory.

| USER_A | SERVER_S | USER_B |
|---|---|---|
| Ka | <Ea,Ka> | Kb |
| Es | Ks | Es |
| ADD ($10) | Kb_Public | ADD ($10) |
| Ka_Public | Ka | Kb_Public |
|  | Ka_Public |  |

The dynamic procedure on Server_S decrypts Sig_U from Ea using Ka.

| USER_A | SERVER_S | USER_B |
|---|---|---|
| Ka | <Ea,Ka,Sig_U> | Kb |
| Es | Ks | Es |
| ADD ($10) | Kb_Public | ADD ($10) |
| Ka_Public | Ka | Kb_Public |
|  | Ka_Public |  |

The dynamic procedure on Server_S loads the value of Kb_Public from server memory.

| USER_A | SERVER_S | USER_B |
|---|---|---|
| Ka | <Ea,Ka,Sig_U,Kb_Public> | Kb |
| Es | Ks | Es |
| ADD ($10) | Kb_Public | ADD ($10) |
| Ka_Public | Ka | Kb_Public |
|  | Ka_Public |  |

The dynamic procedure on Server_S encrypts Eb from Sig_U using Kb_Public.

| USER_A | SERVER_S | USER_B |
|---|---|---|
| Ka | <Ea,Ka,Sig_U,Kb_Public,Eb> | Kb |
| Es | Ks | Es |
| ADD ($10) | Kb_Public | ADD ($10) |
| Ka_Public | Ka | Kb_Public |
|  | Ka_Public |  |

The dynamic procedure on Server_S stores Eb within a self-destructive database.

| USER_A | SERVER_S | USER_B |
|---|---|---|
| Ka | <Ea,Ka,Sig_U,Kb_Public,Eb> -- [Eb] | Kb |
| Es | Ks | Es |
| ADD ($10) | Kb_Public | ADD ($10) |
| Ka_Public | Ka | Kb_Public |
|  | Ka_Public |  |

The dynamic procedure is finally terminated. User_A and User_B are both notified that the transaction has been successfully fully completed.

| USER_A | SERVER_S | USER_B |
|---|---|---|
| Ka | [Eb] | Kb |
| Es | Ks | Es |
| ADD ($10) | Kb_Public | ADD ($10) |
| Ka_Public | Ka | Kb_Public |
|  | Ka_Public |  |

The final result of the transaction is a newly-generated CryptoSquare for User_B.

| USER_A | SERVER_S | USER_B |
|---|---|---|
| Ka | [Eb] | Kb |
| Es | Ks | Es |
| ADD ($10) | Kb_Public | ADD ($10) |
| Ka_Public | Ka | Kb_Public |
|  | Ka_Public |  |

User_B, in tandem with the Server, now controls the resources associated with address ADD. Furthermore, User_B, as well as User_A and Server_S, all remain completely unaware of the ADD-associated signatures. Finally, User_A lacks the means to obtain both signature keys without direct permission from User_B.

It appears that CryptoSquare-based transformations allow us to efficiently transfer funds between two users without relying on the BlockChain. We may geometrically display these transformations by visualizing the movements of cryptographic variables across the eight corners of a cube. The two transforming CryptoSquares for Users A and B are present on two faces of this three-dimensional "CryptoCube." Thusly, we shall refer to all CyptoSquare-dependent Off-Chain transactions as "CryptoCubic Transactions." These CryptoCubic transactions form the core of our CryptoCubic Protocol.

**5 AUTHENTICATION IN THE CRYPTOCUBIC PROTOCAL**

The following additions to our CryptoCubic Protocol will guarantee direct authentication Between User_A, Server_S, and User_B. Let us consider the early stage of the initial relationship between User_A and Server_S, when the two MutiSig signatures are first generated.

| USER_A | SERVER_S |
|---|---|
| Ka | <Ks,Ka_Public,Sig_U,Sig_S,ADD> |
| Ka_Public | Ks |
|  | Ka_Public |

Server_S stores an SHA-outputted Hash of Sig_S within its memory contents.

| USER_A | SERVER_S |
|---|---|
| Ka | <Ks,Ka_Public,Sig_U,Sig_S,ADD> |
| Ka_Public | Ks |
|  | Ka_Public |
|  |  |
|  | Hash |

The Hash remains within the Server's memory after the CryptoSquare relationship is established.

| USER_A | SERVER_S |
|---|---|
| Ka | [Ea] |
| Es | Ks |
| ADD | Ka_Public |
| Ka_Public | |
| | |
| | Hash |

Later, User_A and User_B initialize a CryptoCubic transaction.

| USER_A | SERVER_S | USER_B |
|---|---|---|
| Ka | [Ea] | Kb |
| Es | Ks | Es |
| ADD ($10) | Ka_Public | Kb_Public |
| Ka_Public | | ADD ($10) |
| | | |
| | Hash | |

At this point in the transaction, User_B transfers Kb_Public to User_A.

| USER_A | SERVER_S | USER_B |
|---|---|---|
| Ka | [Ea] | Kb |
| Es | Ks | Es |
| ADD ($10) | Ka_Public | Kb_Public |
| Ka_Public | | ADD ($10) |
| | | |
| Kb_Public | Hash | |

Afterwards, User_A transfers Kb_Public to Server_S.

| USER_A | SERVER_S | USER_B |
|---|---|---|
| Ka | <Ea> | Kb |
| Es | Ks | Es |
| ADD ($10) | Kb_Public | ADD ($10) |
| Ka_Public | Ka | Kb_Public |
| | Ka_Public | |
| | | |
| Kb_Public | Hash | |

Server_S must now authenticate the true identity of User_A. The Server does so by creating a randomized token-string; Token_A.

| USER_A | SERVER_S | USER_B |
|---|---|---|
| Ka | <Ea> | Kb |
| Es | Ks | Es |
| ADD ($10) | Kb_Public | ADD ($10) |
| Ka_Public | Ka | Kb_Public |
|  | Ka_Public |  |
|  |  |  |
| Kb_Public | Hash |  |
|  | Token_A |  |

The Server encrypts the contents of Token_A with Ka_Public, thereby producing cypher Et_A.

| USER_A | SERVER_S | USER_B |
|---|---|---|
| Ka | <Ea> | Kb |
| Es | Ks | Es |
| ADD ($10) | Kb_Public | ADD ($10) |
| Ka_Public | Ka | Kb_Public |
|  | Ka_Public |  |
|  |  |  |
| Kb_Public | Hash |  |
|  | Token_A |  |
|  | Et_A |  |

Et_A is transferred back to User_A. User_A decrypts it; outputting the variable Token_A2. Token_A2 is transferred to the Server. Server_S confirms that Token_A is equivalent to Token_A2. User_A authentication is now complete, and the transaction may continue.

| USER_A | SERVER_S | USER_B |
|---|---|---|
| Ka | <Ea> | Kb |
| Es | Ks | Es |
| ADD ($10) | Kb_Public | ADD ($10) |
| Ka_Public | Ka | Kb_Public |
|  | Ka_Public |  |
|  |  |  |
| Kb_Public | Hash |  |
| Et_A | Token_A |  |
| Token_A2 | Et_A |  |
|  | Token_A2 |  |

Server_S makes preparations to authenticate the identity of User_B. It does so by creating Token_B and the cypher Et_B.

| USER_A | SERVER_S | USER_B |
|---|---|---|
| Ka | <Ea> | Kb |
| Es | Ks | Es |
| ADD ($10) | Kb_Public | ADD ($10) |
| Ka_Public | Ka | Kb_Public |
|  | Ka_Public |  |
|  |  |  |
| Kb_Public | Hash |  |
| Et_A | Token_A |  |
| Token_A2 | Et_A |  |
|  | Token_A2 |  |
|  | Token_B |  |
|  | Et_B |  |

Afterwards, the Server receives a contact request from User_B. Server_S establishes the identity of User_B using the aforementioned token-exchange schema.

| USER_A | SERVER_S | USER_B |
|---|---|---|
| Ka | <Ea> | Kb |
| Es | Ks | Es |
| ADD ($10) | Kb_Public | ADD ($10) |
| Ka_Public | Ka | Kb_Public |
|  | Ka_Public |  |
|  |  |  |
| Kb_Public | Hash |  |
| Et_A | Token_A |  |
| Token_A2 | Et_A |  |
|  | Token_A2 |  |
|  | Token_B | Et_B |
|  | Et_B | Token_B2 |
|  | Token_B2 |  |

Server_S proceeds to transfer the value of Hash to User_B.

| USER_A | SERVER_S | USER_B |
|---|---|---|
| Ka | <Ea> | Kb |
| Es | Ks | Es |
| ADD ($10) | Kb_Public | ADD ($10) |
| Ka_Public | Ka | Kb_Public |
|  | Ka_Public |  |
|  |  |  |
| Kb_Public | Hash | Hash |
| Et_A | Token_A |  |
| Token_A2 | Et_A |  |
|  | Token_A2 |  |
|  | Token_B | Et_B |
|  | Et_B | Token_B2 |
|  | Token_B2 |  |

User_B executes an SHA-hash on cypher Es, outputting the variable Hash2.

| USER_A | SERVER_S | USER_B |
|---|---|---|
| Ka | <Ea> | Kb |
| Es | Ks | Es |
| ADD ($10) | Kb_Public | ADD ($10) |
| Ka_Public | Ka | Kb_Public |
|  | Ka_Public |  |
|  |  |  |
| Kb_Public | Hash | Hash |
| Et_A | Token_A | Hash2 |
| Token_A2 | Et_A |  |
|  | Token_A2 |  |
|  | Token_B | Et_B |
|  | Et_B | Token_B2 |
|  | Token_B2 |  |

User_B authenticates that Hash is identical to Hash2. This confirms that User_A has transferred over a non-counterfeit Es.

| USER_A | SERVER_S | USER_B |
|---|---|---|
| Ka | <Ea> | Kb |
| Es | Ks | Es |
| ADD ($10) | Kb_Public | ADD ($10) |
| Ka_Public | Ka | Kb_Public |
|  | Ka_Public |  |
|  |  |  |
| Kb_Public | Hash | Hash |
| Et_A | Token_A | Hash2 |
| Token_A2 | Et_A |  |
|  | Token_A2 |  |
|  | Token_B | Et_B |
|  | Et_B | Token_B2 |
|  | Token_B2 |  |

Authentication is officially completed. The CryptoCubic transaction proceeds as specified; leading to the creation of a CryptoSquare associated with User_B.

| USER_A | SERVER_S | USER_B |
|---|---|---|
| Ka | [Eb] | Kb |
| Es | Ks | Es |
| ADD ($10) | Kb_Public | ADD ($10) |
| Ka_Public | Ka | Kb_Public |
|  | Ka_Public |  |
|  |  |  |
| Kb_Public | Hash | Hash |
| Et_A | Token_A | Hash2 |
| Token_A2 | Et_A |  |
|  | Token_A2 |  |
|  | Token_B | Et_B |
|  | Et_B | Token_B2 |
|  | Token_B2 |  |